\documentstyle[multicol,tighten,aps,epsfig]{revtex} 
\begin{document}
\draft

\def\AB{$A+B\to 0$}
\def\av#1{\langle#1\rangle}
\def\lab{\ell_{\rm AB}}
\def\etal{{\it et al.}}
\def\a{\alpha}
\def\Noc{N_{\rm oc}}
\def\lsim{\mathrel{\raise.34ex\hbox{$<$\kern-.8em\lower1ex\hbox{$\sim$}}}}

\title{Large Scale Simulations of Two-Species Annihilation, \AB, with Drift}
\author{Yinon Shafrir and Daniel ben-Avraham\footnote
  {{\bf e-mail:} benavraham@clarkson.edu}}

\address{Physics Department, and Clarkson Institute for Statistical
Physics (CISP), \\ Clarkson University, Potsdam, NY 13699-5820}
\maketitle

\begin{abstract} 
We present results of computer simulations of the diffusion-limited reaction
process \AB, on the line, under extreme drift conditions, for lattices of up
to $2^{27}$ sites, and where the process proceeds to completion (no
particles left).  These enormous simulations are made possible by the
renormalized reaction-cell method (RRC). Our results allow us to resolve an
existing controversy about the rate of growth of domain sizes, and about
corrections to scaling of the concentration decay.
\end{abstract}
\begin{multicols}{2}

\section{introduction}
Diffusion-limited reactions have attracted much interest in recent
years~\cite{reactions,kang85}.  The kinetics of such systems is dominated
by local fluctuations in the concentration of the reactants, thus posing a
formidable problem which has not yet been solved: there
exists no comprehensive theoretical approach for the analysis of
diffusion-limited processes.  In this state of affairs, computer simulations
provide a much needed input.

A basic example is diffusion-limited two-species annihilation,
\AB~\cite{kang85}. Particles of two different species, $A$ and $B$, diffuse
and react immediately, and irreversibly, upon encounter.  The product is
assumed to be an inert species which does not affect the process in any
significant way. When the two species diffuse with the same diffusion
constant, and their initial concentrations are equal, the concentration (of
either species) decays anomalously slow: $c\sim 1/t^{d/4}$, for space
dimension $d<4$.  In
$d\geq 4$, local fluctuations cease to dominate and one obtains the
``mean-field" result of the reaction-limited case, $c\sim 1/t$. 
In low dimensions, the process is slowed down because of the
formation of domains consisting of particles of only one of the
species.  These domains grow with time, and various length scales
characteristic of their geometry are then of interest.  Surprisingly, when a
drift field is imposed, i.e., when the particles hop preferentially in a
certain direction, the kinetics is altered dramatically~\cite{remark}.  For
example, in one dimension one finds that $c\sim 1/t^{1/3}$, instead of
$c\sim 1/t^{1/4}$ without drift~\cite{janowsky95,ispolatov95}.  The anomalous
behavior, and the difference between drift and no drift is most pronounced in
$d=1$, and we shall henceforth restrict ourselves to this case.

The kinetics of two-species annihilation with drift is not
as well understood as without drift.  While everybody agrees on the long
time asymptotic decay of the concentration,
$c\sim1/t^{1/3}$~\cite{janowsky95,ispolatov95,benavraham95,park00}, other
behavior is controversial:  Ispolatov
\etal,~\cite{ispolatov95} propose that the typical domain size grows as
$\ell\sim t^{\a}$ with
$\a=2/3$, and that there is a correction to the concentration decay, of the
form $c\sim t^{-1/3}(A+Bt^{-x})$, with $x=1/6$ ($A$ and $B$ are constants). 
On the other hand, Janowsky~\cite{janowsky95} argues that the domain growth
exponent is
$\a=7/12$. Comparing this with the superdiffusive length scale
$\ell\sim t^{2/3}$, which naturally arises in diffusion with
drift and hard core interaction~\cite{vanbeijeren85,gwa92}, one
obtains the correction exponent $x=1/12$.  Janowsky also predicts that the
typical distance between domains grows as $\lab\sim t^{3/8}$, similar to the
process without diffusion.  These small disagreements could not be resolved
by the computer simulation data presented at the time.  Here we report on
large scale simulations of \AB\ with drift, using the method of renormalized
reaction-cells (RRC)~\cite{benavraham87,dawkins00}.  This technique has
enabled us to simulate the process in one-dimensional lattices of up to
$2^{27}\approx 1.34\times10^8$ sites, to completion, until there are no
particles left.  The results confirm the concentration decay of
$c\sim1/t^{1/3}$, and strongly support the exponents and
corrections to scaling predicted by Janowsky.

\section{the RRC algorithm}
The two-species annihilation process is modeled as follows.  The sites of
a one-dimensional lattice are either empty or occupied by a  
single $A$ or $B$ particle.  (Periodic boundary conditions are
imposed, so the lattice is effectively a ring.)  The particles undergo
diffusion under extreme drift conditions: at each Monte Carlo step a particle
is chosen randomly and is moved to the nearest site to its right.  Diffusion
results because of the random order of the updates.  If the target site is
occupied by a particle of the opposite species, then both particles are
removed from the system, mimicking the reaction \AB.  On the other hand, if
the target site is occupied by a particle of the same species, then the move
is disallowed and it does not take place.  This implements the excluded
volume interaction between like particles, without which drift has no
significant effect on the kinetics.  Regardless of the outcome, time is
incremented by $1/N(t)$, where $N(t)$ is the total number of particles
remaining in the system.  Thus, in one unit time, all of the particles
move (or attempt to move) once, on average.  

As the simulation proceeds, the particle concentration declines and the
typical distance between particles increases.  The time spent on simulating
the diffusive motion of the particles until they interact grows even
faster, as the square of the distance between them.  Because of that, computer
simulations are limited to relatively short times.  This problem is overcome
by the RRC method~\cite{benavraham87,dawkins00}, which we have adapted for the
case of \AB\ with drift.

In our implementation of the RRC method the particles occupy cells, rather
than sites.  Each cell may be either empty, or occupied by one or more
particles of only one species.  Whenever the overall concentration drops by
half, the lattice undergoes a renormalization process. Every two adjacent
cells are merged into a new cell: cells $2k$ and
$(2k+1)$ in the old lattice are combined into a new cell, $k'$, in the new
lattice. The contents of cell $k'$ is the combined contents of the two parent
cells, $2k$ and $(2k+1)$: Suppose that the number of particles in the
parent cells is $n_{2k}$ and $n_{2k+1}$, then, if the particles in both cells
are of the same species, there would be
$n_{2k}+n_{2k+1}$ particles of that same species in $k'$.  Otherwise, the
annihilation reaction takes effect and $k'$ contains
$|n_{2k}-n_{2k+1}|$ particles of that species that was initially in the
majority.  The particles hop from one cell to the next in a single Monte
Carlo step, and because the cells keep growing the diffusion process is
greatly sped up:  In order to diffuse out of a renormalized
cell twice as large as that of the previous generation, a particle requires a
time longer by a factor of four.  Thus, physical time is simulated faster
with each renormalization step.

A Monte Carlo step consists of selecting a {\it non-empty\/} cell at random
and moving a particle from within it to the adjacent cell to its right.  The
move is allowed only if the target cell is empty, or if it contains
particles of the opposite species.  The number of particles in the cell of
origin decreases by one; the number of particles in the target cell increases
by one, if it was initially empty, or decreases by one, if it
contained particles of the opposite species.  Note, however, the following
objections:  (1)~Different cells contain different numbers of
particles, so the more populated cells should be selected more often.  (2)~The
excluded volume interaction seems not to be fully accounted for, since cells
may contain more than one particle.  We will now see how both problems
may be elegantly addressed, by carefully tuning the time update associated
with each Monte Carlo step.

Let $m(t)$ be the maximum number of particles present at any given cell in the
system at time $t$.  To make the total rate of all possible processes
within each non-empty cell equal, assume that cells with $n<m$ particles
contain also
$m-n$ ``ghost'' particles.  The ghost particles require the same time to hop
(or to attempt a hop) just as real particles do, but their presence has no
other effect.  Each occupied cell now contains a total of $m$ particles and
the total rates are equal. The time increment associated with a Monte
Carlo step should be $\Delta t = 4^k/m\Noc$, where $k$ is the number of
renormalizations since the start of the run, and $\Noc(t)$ is the current
number of occupied cells.  To take the excluded volume effect within a
cell into account, we assume that only the rightmost (real) particle may
attempt a hop to the adjacent cell on the right, for only it is unimpeded by
the presence of the other (real) particles in the cell.  But the probability
that the rightmost particle is selected for the attempt is $1/m$.  Instead, we
can assume that the rightmost particle has been selected, with probability
one, but that a longer time has elapsed, on average, by a factor of $m$:
$\Delta t= m(4^k/m\Noc)=4^k/\Noc$.  Algorithmically, then, there is no need
to worry about ghost particles, nor about the actual value of
$m$.  One simply selects one of the $\Noc$ cells at random and executes the
Monte Carlo step, then increments time by $\Delta t = 4^k/\Noc$.

As a final observation we note that because the concentration keeps roughly
constant throughout the simulation, it may be desirable to work with low
initial concentrations. Otherwise, particles are typically separated by but a
few cells, and one should worry about the effect of such coarseness on the
results.  Also, a low concentration reduces the fraction of cells with more
than one particle, thus minimizing concerns about the approximate treatment
of the excluded volume interaction within multiply occupied cells.  We have
found satisfactory results with initial concentrations of $1/16$ (of the
two species combined), and we have used this value for all of the simulation
results reported below. 

\section{Results}
Our first goal is to convince ourselves of the reliability of the RRC method.
To this end, we have simulated the \AB\ process on lattices of $2^{16}=65,536$
sites, in both the RRC and the traditional simulation method.  These lattices
are small enough to enable the simulation of the process by the traditional
method to completion, thus providing a benchmark test throughout all its three
stages: (1)~the initial phase, until particles diffuse across the distance
separating each other and start reacting in mass (this takes of the order of
$1/c(0)^2$), (2)~the main phase, characterized by the  $1/t^{1/3}$ decay, and
typical of infinite systems, and (3)~the end phase, where finite-size effects
kick in and the decay speeds up to exponential rate.  On the other hand, the
system is large enough to let us examine the effect of the renormalizations:
with $2^{16}$ sites and $c(0)=1/16$ the RRC method requires $12$
renormalizations.  In Fig.~1 we compare the particle concentration as
obtained by the two methods.  The agreement is very good, and there are no
discernible anomalies associated with the renormalizations.  The agreement
is also as good for other quantities measured, such as the domain size and
the distance between domains.

Having gained some confidence in the RRC method, we proceeded to perform
larger simulations.  In Fig.~2a we show the surviving number of particles,
$N(t)$, at time $t$, for a system of $2^{27}\approx 1.34\times10^8$ sites. 
The initial number of particles is close to $10^7$, and the process has been
simulated to completion, after about $t=3\times10^{15}$ time units.  (Recall
that in one unit of time all the particles in the system attempt one move, on
average.)  In comparison, other simulations to
date~\cite{janowsky95,ispolatov95,benavraham95} have been carried on lattices
of
$5\times10^5$ to $4\times10^6$ sites, and up to 
$t=2\times10^5$ to $10^6$ time units.  The three stages of the process are
clearly visible in Fig.~2a, but here the main phase alone spans nearly ten
orders of magnitude. In Fig.~2b we show the local slopes of the plot in
part~(a).  It can be seen that the approach to the predicted value, $1/3$, is
extremely slow.  Corrections to scaling are sizable even after very long
times.  We return to the issue of corrections later on.

Next we look at the inter-domain distance, $\lab(t)$ --- the distance between
the last particle in a domain and the first particle in the domain next to it.
The growth predicted by Janowsky, of $\lab\sim t^{3/8}$, seems to hold true
throughout the main phase.  Towards the end of the process, the inter-domain
distance grows somewhat faster (Fig.~3).

The first quantity under dispute is the scaling of the domain length.  In
Fig.~4 we show a log-log plot of the average domain length, $\ell(t)$, as a
function of time.  Slopes of $2/3$ (Ispolatov \etal) and $7/12$ (Janowsky)
are shown for comparison.  While the results of this figure seem to favor the
scaling proposed by Janowsky, they are not convincingly conclusive.  For
additional evidence, we turn to the following simple-minded scaling
analysis.  We argue that the particle concentration scales as
\begin{equation}
c(t)=L^{-1/3\a}\rho(t/L^{1/\a})\;,
\end{equation}
where $\a$ is the domain growth exponent, $\ell(t)\sim t^{\a}$, and $L$ is
the lattice size.  We simply assume that the typical domain size is the
fastest growing length scale in the system.  (This is in agreement with all
theoretical predictions.)  The proposed scaling form reflects our
expectation that the concentration decay throughout the main and the final
phases of the process be a function of $\ell/L\sim(t/L^{1/\a})^{\a}$.  On the
other hand, for infinitely large lattices the
$L$ dependence ought to disappear and one expects the pure power-law
behavior
$c(t)\sim t^{-1/3}$.  This can only be if $\rho(z)\sim z^{-1/3}$, for $z\to0$,
and the prefactor of $L^{-1/3\a}$ is necessary to eliminate the
dependence on $L$.  In Fig.~5, we plot $L^{1/3\a}c(t)$ as a function of
$t/L^{1/\a}$ for different lattice sizes and for $\a=2/3$ and $7/12$. 
The data collapse, in particular near the end of the process, is clearly
better for $\a=7/12$. 

Finally, let us address the issue of corrections to scaling of the
concentration decay.  According to Ispolatov \etal,~\cite{ispolatov95} the
corrections are of the form
\begin{equation}
c(t)\sim t^{-1/3}(A+Bt^{-x})\;,
\end{equation}
where $A$ and $B$ are constants, and the correction exponent is $x=1/6$.  This
results from a comparison of the domain growth as predicted by an inviscid
($\ell\sim t^{2/3}$) vs.~a viscid ($\ell\sim t^{1/2}$) Burgers equation. 
According to Janowsky~\cite{janowsky95}, the superdiffusive length $t^{2/3}$
still plays the principal role in determining the concentration decay,
however, in his case the predicted domain growth of $\ell\sim t^{7/12}$
provides a stronger correction than the viscid limit considered by Ispolatov
\etal.  Hence, one should expect
$x=1/12$ rather than $1/6$.

Our strategy consists of performing a least squares linear fit of $A+Bt^{-x}$
to $t^{1/3}c(t)$, for different powers $x$, and searching for the value of $x$
which minimizes the error.  The asymptotic form~(2) is expected to work only
in the main phase, and the sticky part of our procedure is deciding which
times demarcate this region.  By choosing different regions between
$t=10^3$ and $t=10^{12}$ in the data for $2^{27}$-site lattices, we conclude
that $1/16\lsim x\lsim 1/8$, in good agreement with $x=1/12$, and excluding,
quite confidently, the possibility of $x=1/6$. In Fig.~6a we show best fits
for the region $t=10^3$ -- $10^8$, where our data is most reliable.  It can
be clearly seen that $x=1/12$ provides a much better fit than $x=1/6$.  In
Fig.~6b we show how the same parameters (for $x=1/6$) compare to the data at
all times.  The fit is reasonably good throughout all of the main phase.

\section{Summary and discussion}
We have presented large scale simulation results of diffusion-limited
two-species annihilation, \AB, with drift, in one dimension, using the RRC
method.  Our simulations are large enough to resolve an existing dispute
about the rate of domain growth, and about the corrections to scaling of the
concentration decay.  Two different length scales are associated with a
diffusing particle subject to drift, and so it is not obvious {\it a priori\/}
that the RRC method should be successful in such cases, for the renormalization
step can account for only one of the two scales.  We conclude that the RRC
method is capable of handling reaction-diffusion systems with drift, and
that the relevant length scale --- the one used in the renormalization step
--- is the diffusive length.   We note that the RRC method is easily
generalized to other reactions and to higher dimensions.  Reactions with
drift  pose a more challenging theoretical problem than reactions without
drift, and we anticipate that computer simulations, including the RRC method,
will continue to be an important tool in their research in years to come.

\acknowledgments
We thank S.~Redner for numerous discussions, and we
gratefully acknowledge support from the National Science Foundation
(PHY-9820569).


\end{multicols} 
\vfil\eject

%
\begin{figure}
\centerline{\epsfxsize=17cm \epsfbox{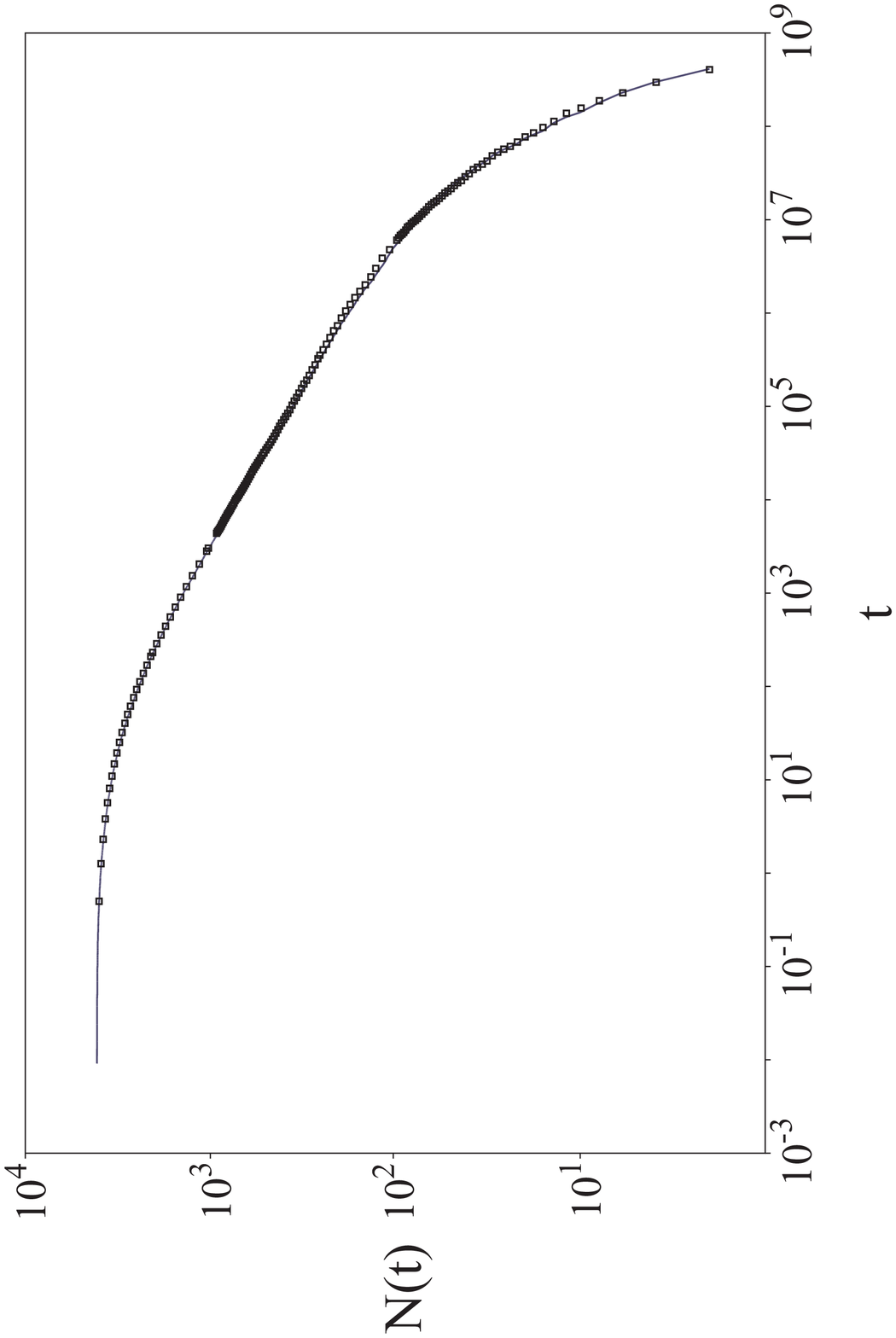}}
\noindent
\centerline{{\bf Fig.~1:} Comparison of cell-indexing results (squares) to RRC
results (solid line) for $2^{16}$-site lattices.}
\end{figure} 
\vfil\eject

\begin{figure}
\centerline{\epsfxsize=17cm \epsfbox{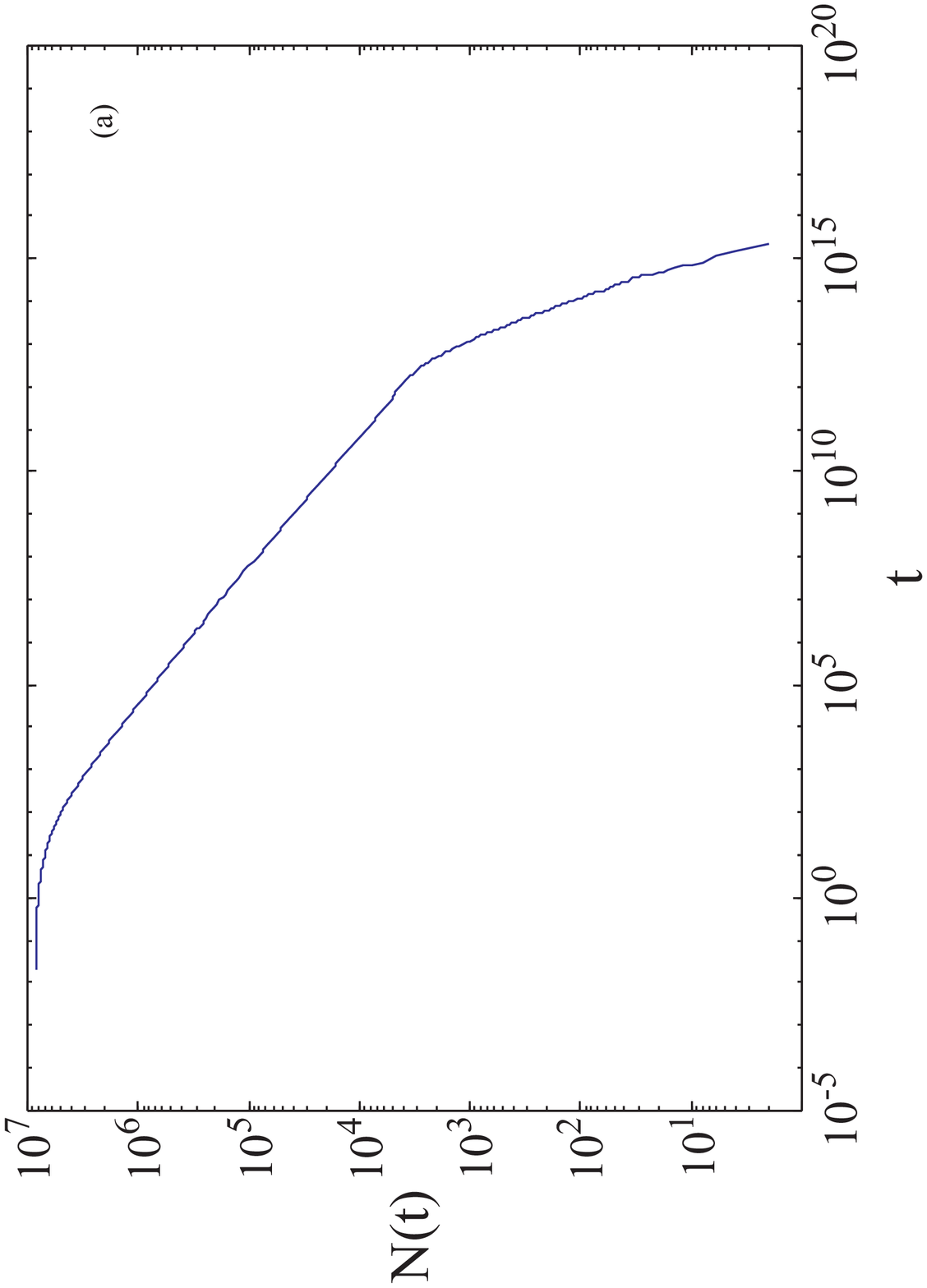}}
\noindent
{{\bf Fig.~2a:} Concentration decay in \AB\ with drift: (a)~Log-log plot of
the number of surviving particles $N(t)$ vs.~$t$.  (b)~Local slopes, $-d\,\ln
N(t)/d\,\ln t$ of the data in (a).  Results are for one run on a 
$2^{27}$-site lattice, and initial concentration $c(0)=1/16$.}
\end{figure} 
\vfil\eject

\begin{figure}
\centerline{\epsfxsize=17cm \epsfbox{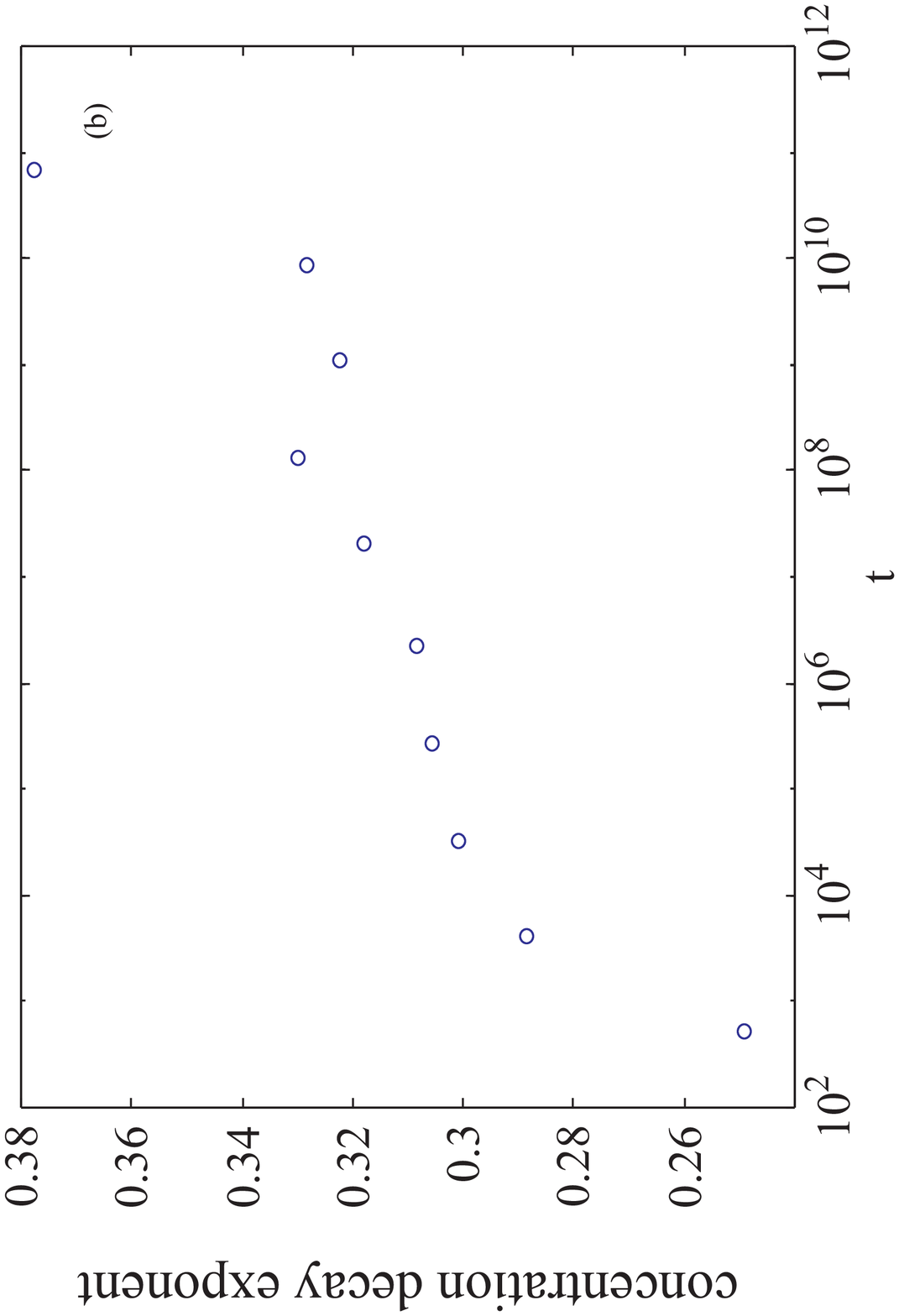}}
\noindent
{{\bf Fig.~2b:} Concentration decay in \AB\ with drift: (a)~Log-log plot of
the number of surviving particles $N(t)$ vs.~$t$.  (b)~Local slopes, $-d\,\ln
N(t)/d\,\ln t$ of the data in (a).  Results are for one run on a 
$2^{27}$-site lattice, and initial concentration $c(0)=1/16$.}
\end{figure} 
\vfil\eject

\begin{figure}
\centerline{\epsfxsize=17cm \epsfbox{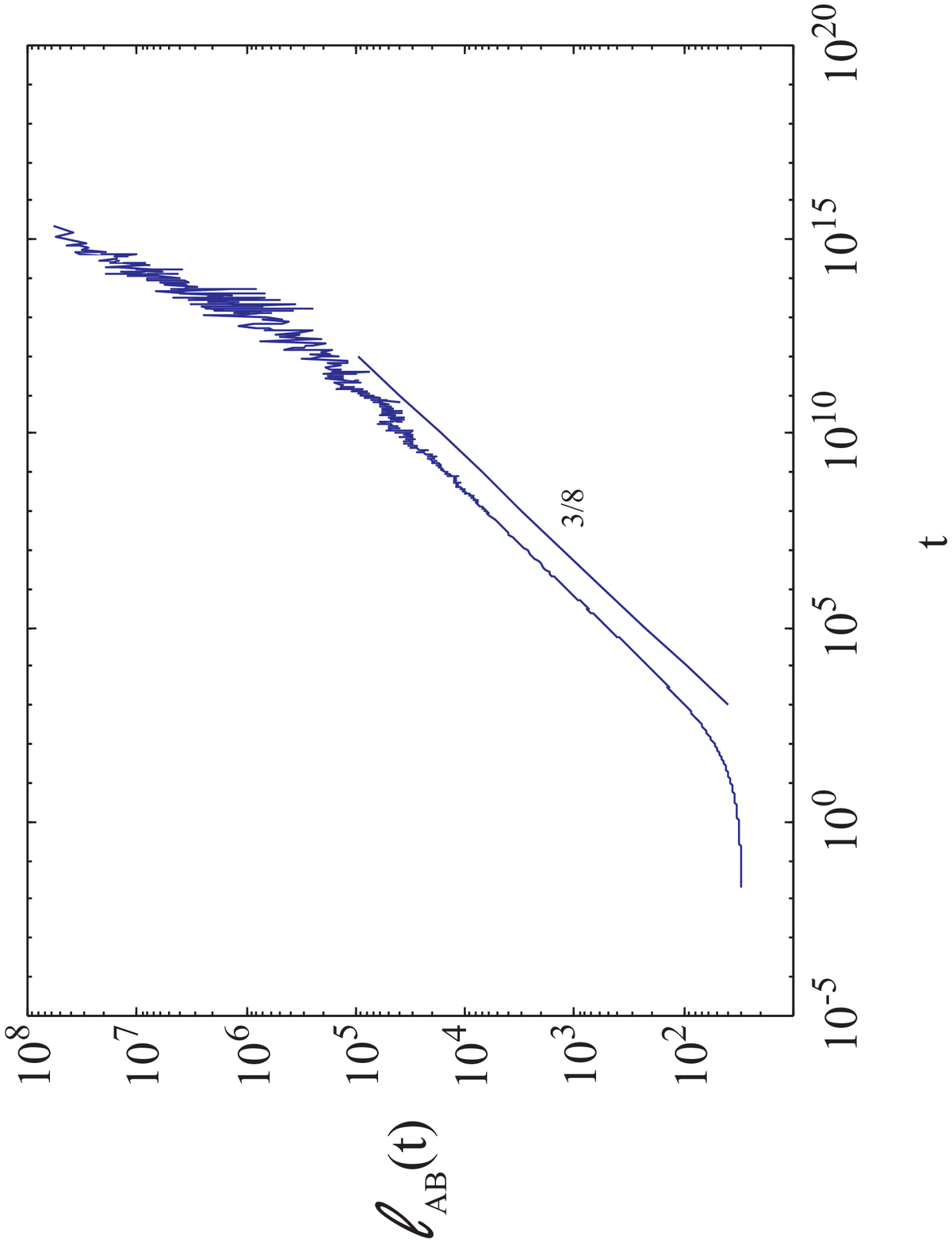}}
\noindent
{{\bf Fig.~3:} Growth of the inter-domain distance, $\lab(t)$, with time, in
the $2^{27}$-site simulation.  The solid line of slope $3/8$ is shown for
comparison.}
\end{figure} 
\vfil\eject

\begin{figure}
\centerline{\epsfxsize=17cm \epsfbox{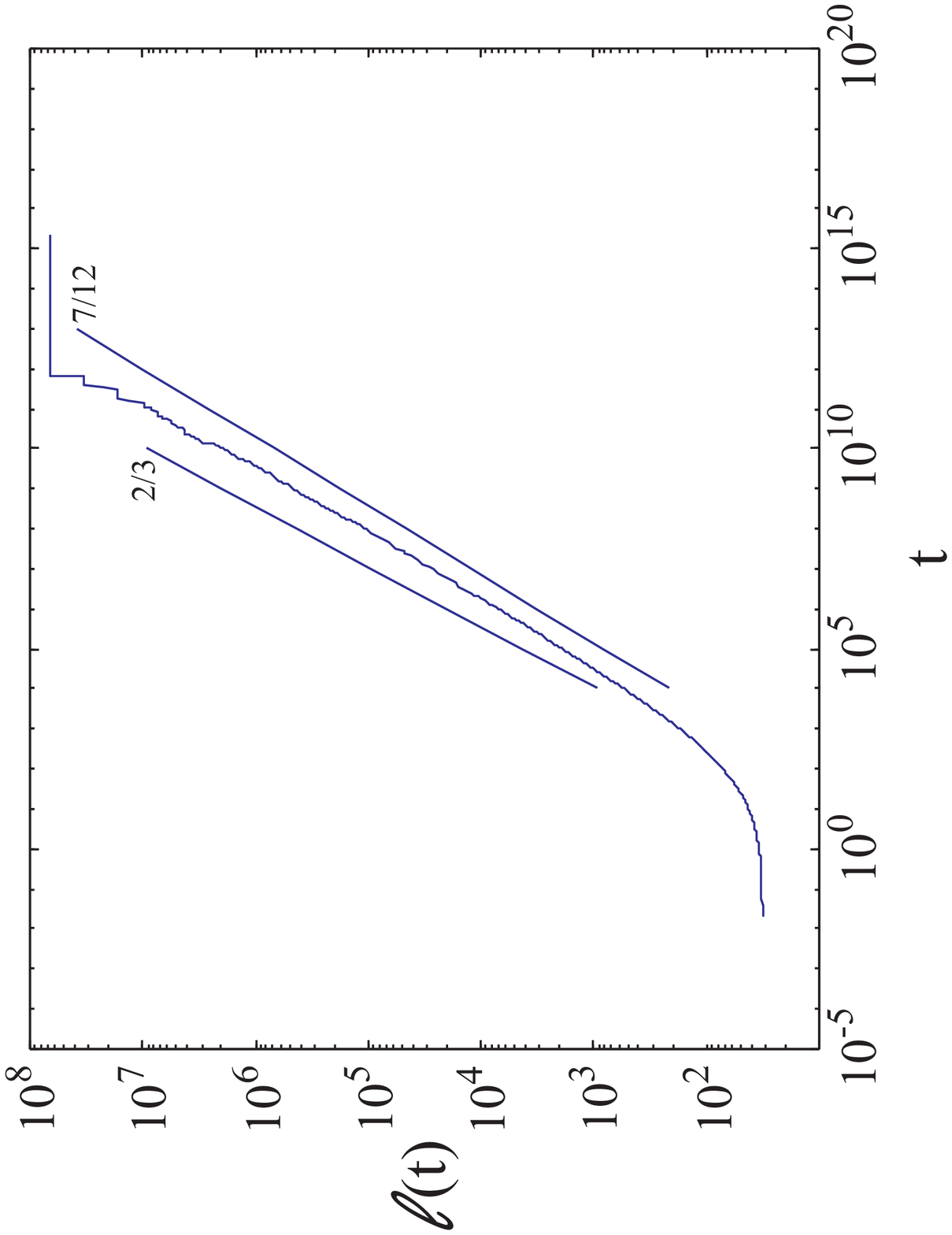}}
\noindent
{{\bf Fig.~4:} Growth of the domain size, $\ell(t)$, with time, in
the $2^{27}$-site simulation.  The solid lines of slope $2/3$ and $7/12$ are
shown for comparison.}
\end{figure} 
\vfil\eject

\begin{figure}
\centerline{\epsfxsize=17cm \epsfbox{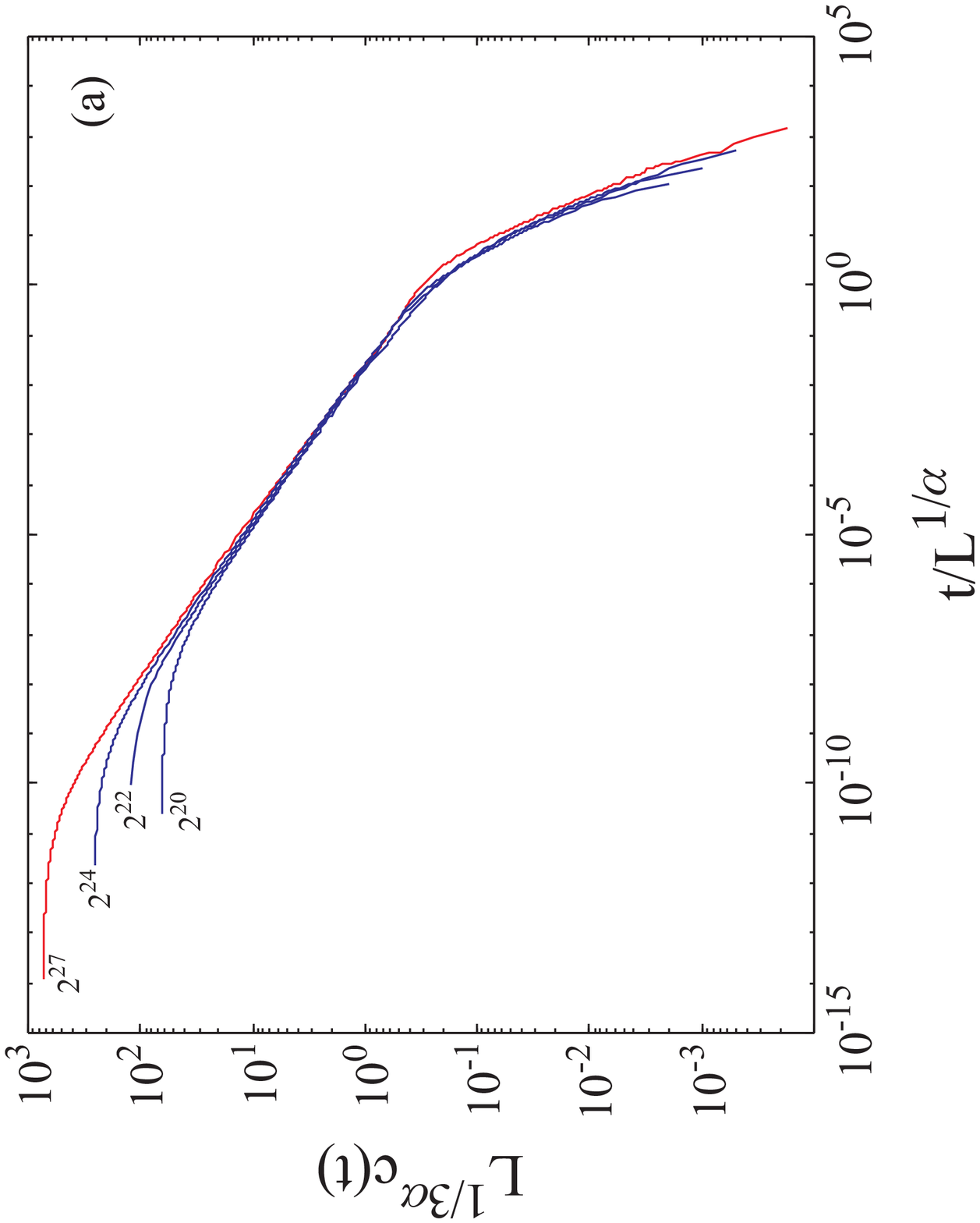}}
\noindent
{{\bf Fig.~5a:} Scaling of the concentration decay.  Plotted is
$L^{1/3\a}c(t)$ vs.~$t/L^{1/\a}$, for simulations on lattices of $L=10^{27}$,
$10^{24}$,
$10^{22}$, and $10^{20}$ sites, for (a)~$\a=2/3$, and (b)~$\a=7/12$.}
\end{figure} 
\vfil\eject

\begin{figure}
\centerline{\epsfxsize=17cm \epsfbox{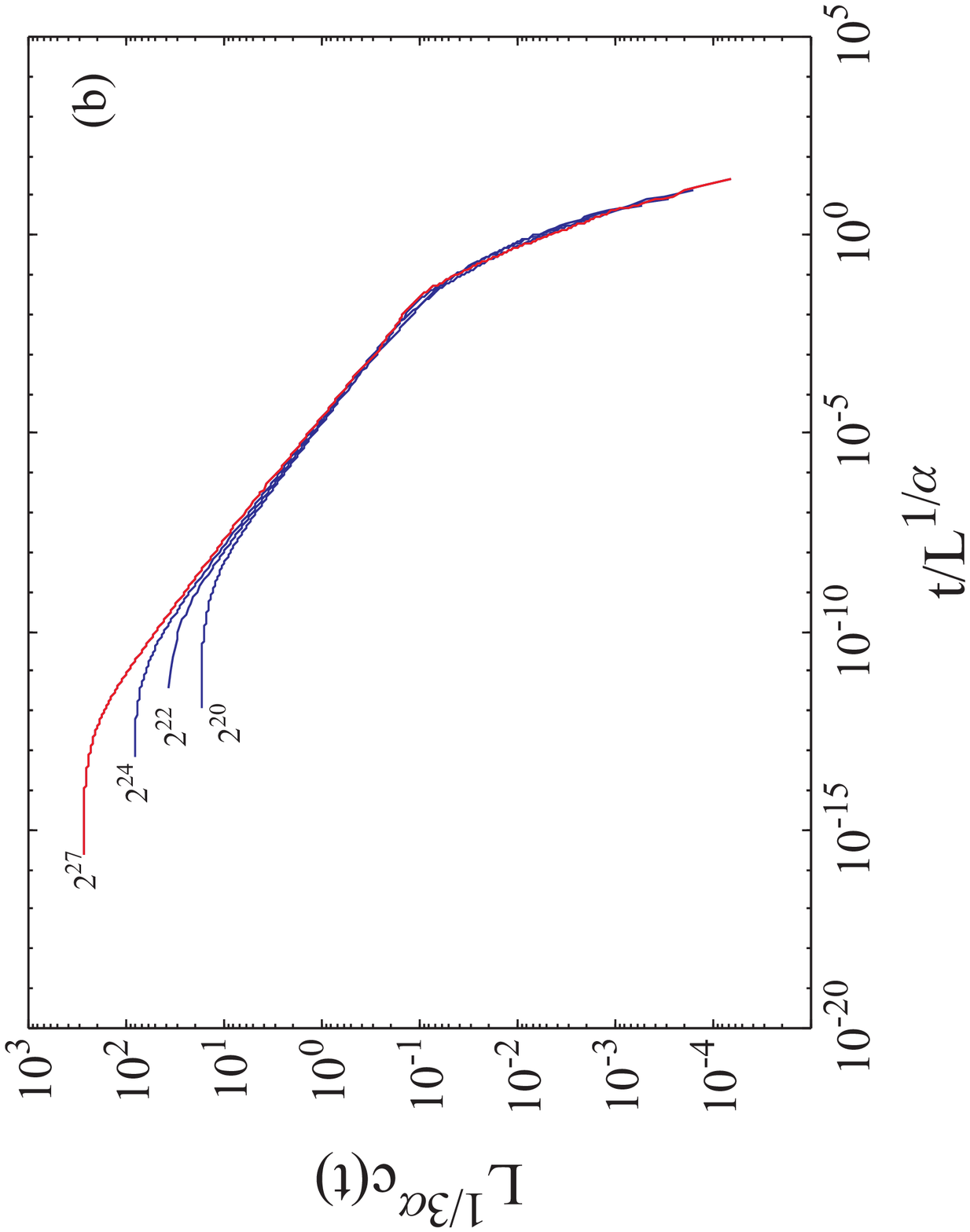}}
\noindent
{{\bf Fig.~5b:} Scaling of the concentration decay.  Plotted is
$L^{1/3\a}c(t)$ vs.~$t/L^{1/\a}$, for simulations on lattices of $L=10^{27}$,
$10^{24}$,
$10^{22}$, and $10^{20}$ sites, for (a)~$\a=2/3$, and (b)~$\a=7/12$.}
\end{figure} 
\vfil\eject

\begin{figure}
\centerline{\epsfxsize=16cm \epsfbox{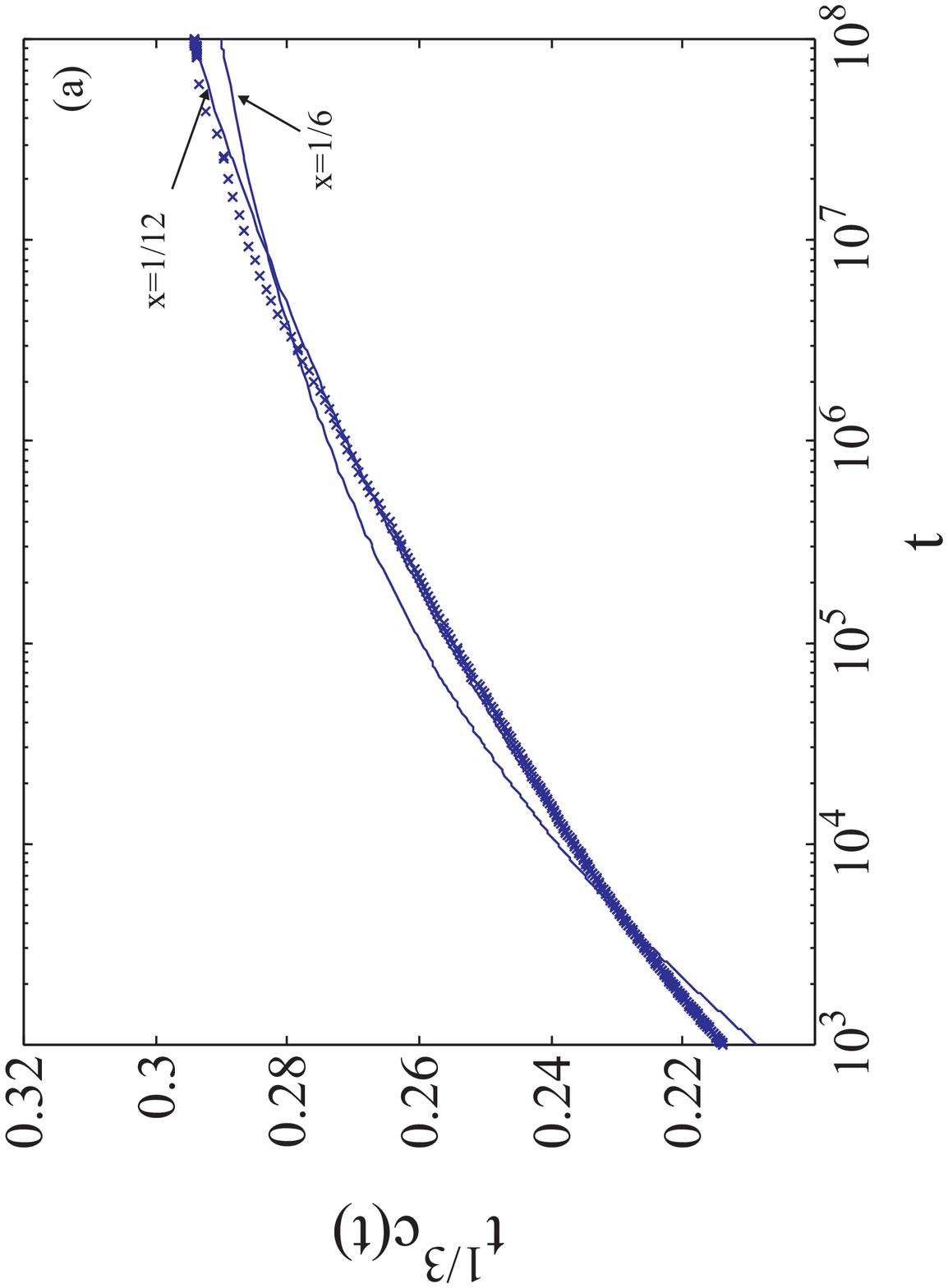}}
\noindent
{{\bf Fig.~6a:} Corrections to scaling of the concentration decay. Plotted is
$t^{1/3}c(t)$ vs.~$t$ for the simulation data on the $2^{27}$-site lattice:
(a)~Best fit of $A+Bt^{-x}$ to the simulation data~($\times$) in the range
$t=10^3$ -- $10^8$, with $x=1/6$ and $x=1/12$ (solid lines).
(b)~Comparison of $A+Bt^{-1/12}$, using the values for $A$ and $B$
found for part~(a), to the simulation data of the complete process.}
\end{figure} 
\vfil\eject

\begin{figure}
\centerline{\epsfxsize=16cm \epsfbox{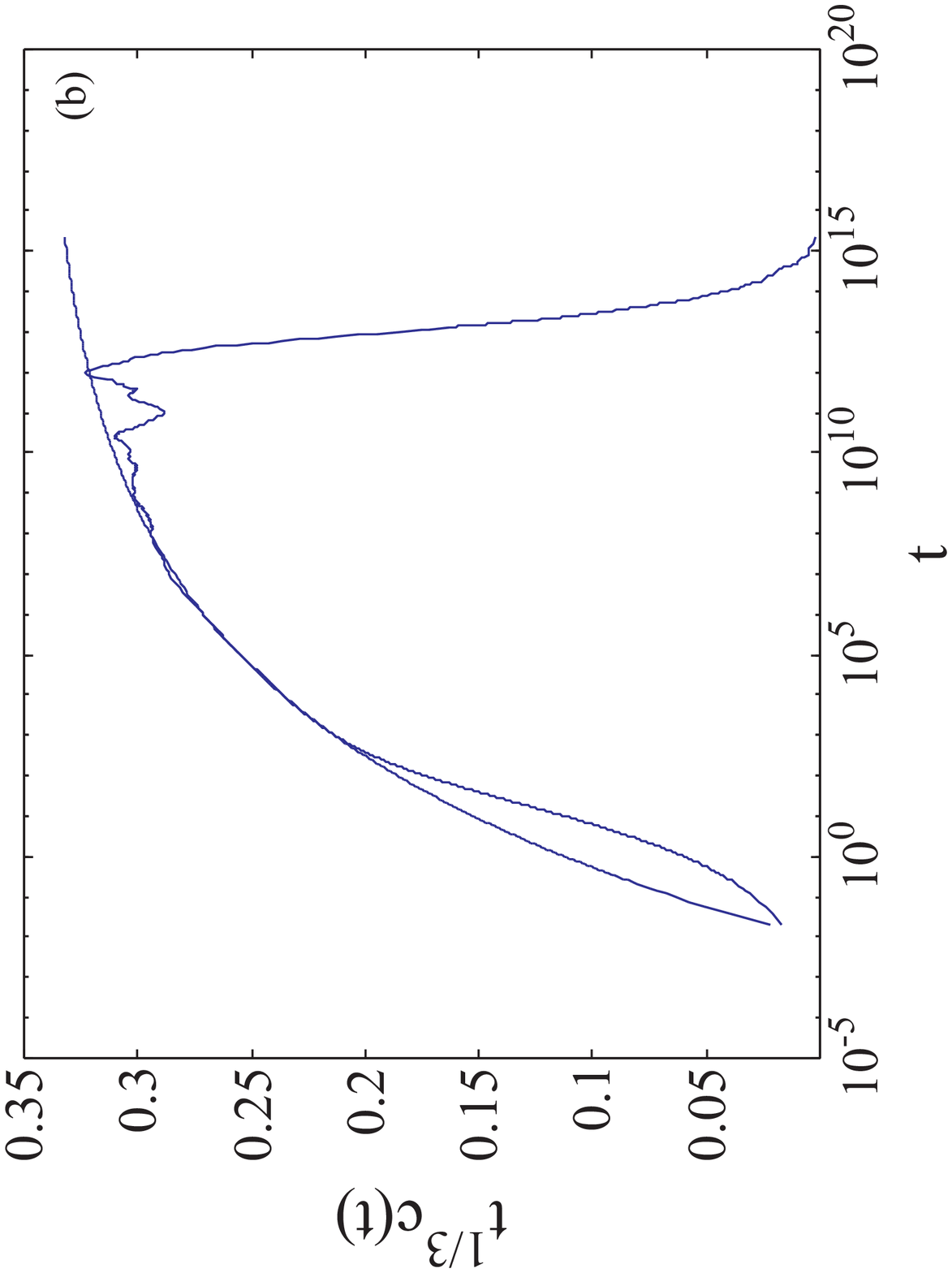}}
\noindent
{{\bf Fig.~6b:} Corrections to scaling of the concentration decay. Plotted is
$t^{1/3}c(t)$ vs.~$t$ for the simulation data on the $2^{27}$-site lattice:
(a)~Best fit of $A+Bt^{-x}$ to the simulation data~($\times$) in the range
$t=10^3$ -- $10^8$, with $x=1/6$ and $x=1/12$ (solid lines).
(b)~Comparison of $A+Bt^{-1/12}$, using the values for $A$ and $B$
found for part~(a), to the simulation data of the complete process.}
\end{figure} 


\end{document}